\documentclass{camera}

\begin{document}

%
\title{Variations on Gravity - Time scales in compact object mergers}

%
\author{William H. Lee}

%
\organization{Instituto de Astronom\'{\i}a, UNAM, Apdo. Postal 70-264, Cd. Universitaria, M\'{e}xico DF 04510, wlee@astroscu.unam.mx}

\maketitle

\begin{abstract}
Compact object mergers, or their collisions in dense stellar environments, involve different time scales. Each of these has its own tempo and dictates a characteristic set of consequences which may, or may not, be observable in what we see as a $\gamma$-ray burst (GRB). I detail here these stages, the state of our knowledge about them, and how they may be important for our understanding of this class of progenitors.
\end{abstract}

%

\section{Spiraling in - {\em Prelude}}


Binaries containing compact objects (neutron stars and black holes) that will merge in less than a Hubble time because of energy and angular momentum losses to gravitational waves are known to exist. These are in fact the most powerful particle accelerators known, where gravity provides the energy leading to the relativistic collision of giant nuclei with $A\approx10^{57}$. The energy  released thus is $\approx \epsilon M c^{2} \approx 6 \times 10^{53} (\epsilon/0.1) (M/3M_{\odot})$~erg, with the bulk being carried by gravitational waves and neutrinos. Some electromagnetic transient is inevitable, and GRBs have been leading candidates for a number of years (see e.g., Lattimer \& Schramm 1976, Paczynski 1986, and reviews by Nakar 2007 and Lee \& Ramirez-Ruiz 2007).

We expect that the in-spiral, when the orbital separation is much larger than the individual stars, will be observed by the advanced gravitational wave detectors (e.g., LIGO, VIRGO). During this phase, the orbital evolution can be computed to high accuracy using post-Newtonian expansions for point masses, and the system parameters (such as stellar masses) extracted from the corresponding observations of its decay.  In general, observation of the gravitational waves emitted allows in principle for the use of such sources as "standard sirens"   for cosmology (Cutler \& Holz 2009). 

The characteristic gravitational signal is that of a "chirp", with the amplitude and frequency rising as the orbit tightens progressively. The typical time scale is essentially the orbital period, and the decay time can be estimated in this regime as $t_{0}/P\approx 10^{5} (P/1{\rm s})^{5/3}$, where $P$ is the current orbital period and $t_{0}$ is the time until merger. For PSR 1913+16 (Hulse \& Taylor 1975) and J0737-3039 (Burgay et al. 2003), $t_{0}\approx 300, 85$~Myr, respectively (the current orbital periods are 8 and 2.4 hours). As the orbit shrinks, the pace quickens until the simple point-mass treatment breaks down completely. 

\section{Merging - {\em Allegro}}

The merger phase, traditionally beginning when the separation is a few stellar radii and extending all the way down to the formation of a single bound object, involves a violent redistribution of mass, energy and angular momentum. Just before the collision, the two members of the pair are rotating at break-up velocities, and form  a dynamically unstable bar-like structure. The instability and subsequent plunge is triggered by a combination of gravitational wave emission and tidal effects (Lai et al. 1993), the latter of which are sensitive to the assumed compressibility at high densities. 

Consider the case of two neutron stars. The mass ratio is close to unity, and the ratio of kinetic to gravitational potential energy exceeds the critical value for dynamical stability, $T/|W| > 0.27$. The system thus responds by ejecting matter through the highest points in the effective potential: the outer Lagrange points $L_2$ and $L_3$, aligned with the bar. A relatively small amount of mass carries away a large fraction of the angular momentum and associated kinetic energy, allowing the two components to come together as a single. At the same time, two large-scale spiral arms are formed from the ejected material. Some of this may have positive total energy and thus be unbound from the system (e.g., Rosswog \& Davies 2002). The remainder will eventually return, with possibly interesting consequences for later evolutionary phases. 

Most of the material ends up in a rapidly rotating configuration, with the total mass 
usually being above the limit for a cold, non-rotating neutron star (Cook et al. 1994). Differential rotation (Baumgarte et al. 2000) may temporarily stabilize this supra-massive neutron star against collapse,  but the ultimate fate and time scale for collapse remains uncertain. 
For a black hole-neutron star encounter (Lee 2000), a large fraction of the stellar material is directly accreted by the black hole. The remainder is left in orbit in a toroidal structure, roughly at the tidal disruption radius. 
The disruption, mass ejection and disk formation time scale is a few orbital periods,  which coincides with the dynamical time for the newly formed bound object. This is $t_{\rm disrupt}\approx {\rm few} \times t_{\rm dyn} \approx {\rm few} \times (G \overline \rho)^{-1/2} \approx 10-20 $~ms. This is usually the span over which such encounters are simulated numerically. 

Tidal disruption and the ensuing evolution transforms a bar-like configuration with a rapidly changing mass quadrupole moment into an azimuthally symmetric structure, so gravitational wave emission ceases promptly at this point. The precise turn-off is a sensitive function of the stellar radius, as it is the key parameter coupling the two components through the tidal field. The cut in the gravitational wave spectrum can then be used to determine the neutron star radius (Faber et al. 2002). With the mass determination from the in-spiral phase, constraints on the mass-radius relationship and the allowed equation of state of dense matter can be placed (Lattimer \& Prakash 2007). This will require observing the gravitational wave emission at frequencies in the kHz range, higher than the standard band for the current LIGO and VIRGO. 

For a wide variety of initial configurations, calculations by various groups using different physics input and numerical schemes appear to converge on the fact that formation of a massive accretion disk, with $M\approx (0.001-0.1) M_{\odot}$ is a robust result (Lee \& Ramirez-Ruiz 2007 and references therein). However,
a number of important open issues remain at this juncture. Among the most pressing are the following: what is the role of magnetic fields? what is the fate of a supra-massive neutron star in case one is formed? on what time scale is a change in its structure likely to be manifested? These have been addressed to varying degrees (see e.g., Price \& Rosswog 2007), but we still lack a definitive answer. 

\section{Accreting - {\em Adagio}}

Once an accretion torus forms, its evolution is subject largely to two driving agents: cooling losses, and transport of angular momentum. In the thermodynamical regime relevant for disks formed from compact mergers ($\rho \approx 10^{11}$~g~cm$^{-3}$, $T \approx 1-10$~MeV), the first comes from neutrino emission, dominated by pair annihilation and $e^{\pm}$ capture onto free nucleons\footnote{This is the so-called hypercritical accretion regime (Chevalier 1989).}. The gas is close to weak and nuclear statistical equilibrium, and may undergo significant neutronization in the process. Photons, as blackbody radiation, are an integral part of the fluid and neutrino optical depths can reach 10-100 for massive disks in the early phase of evolution. A first time scale is then set in such cases by the cooling time, $t_{\rm cool} \approx E_{\rm int}/L_{\nu}$, where $E_{\rm int}$ is the internal energy reservoir in the disk after disruption (recall that it is initially high, as the stars are in virial equilibrium). The associated spans are typically $t_{\rm cool}\approx 0.1$~s, already comparable with the duration of some short GRBs, and much longer than the disruption time scale itself (Lee et al. 2004). 

Any mechanism that transports angular momentum outwards will allow for mass accretion. The Shakura-Sunyaev (1973) $\alpha$-prescription, and the development of the magnetorotational instability (MRI, Balbus \& Hawley 1991) have been studied as providers of this transport, but they are not the only ones relevant. As the disks are fairly massive, fragmentation and self-gravitational instabilities could also play a role. Most modeling to date has been carried out in two dimensions with azimuthal symmetry, allowing for a greater temporal range of study (Lee et al. 2005).  Those that have been carried out in three dimensions (Setiawan et al. 2004) are of limited duration and have not been able to address these issues in greater detail.  This  area clearly deserves further careful study, as we have yet to fully understand and quantify the setting and operation of the equivalent viscous time scale. 

Nevertheless, when the evolution is computed for a range of effective viscosities, with $\alpha=10^{-3}-10^{-1}$, the disks are capable of dissipating and releasing energy efficiently, with $\eta_{\rm acc}=L_{\nu}/\dot{M} c^{2} \simeq 0.1$ (Di Matteo et al. 2002), for periods lasting up to one second (Lee \& Ramirez-Ruiz 2007). Neutrino emission may power part of the GRB itself, but can also be considered as a proxy for central engine activity and accretion, manifesting itself in a relativistic outflow, for example, through magnetically-driven collimated flows (Lyutikov 2006).  Thus the viscous time scale, as a measure of this activity, can in principle be itself one order of magnitude larger than that associated with cooling, and clearly extend to the typical duration scales in short GRBs.

If the central object is not a black hole but a rapidly rotating, highly magnetized neutron star, or magnetar, an entirely different channel for possible activity is available. This can come either purely from the star itself (Usov 1992), or through interaction with the surrounding disk and the generation of neutrino-driven winds (Metzger et al. 2008). As long as the star remains stable to collapse, such systems can account for sustained activity for intervals longer than those associated with GRBs. In either case, the evolution proceeds as well on a time scale at least one order of magnitude longer than that associated with the previous merger phase. 

\section{Coda - {\em Largo}}

Tidal tails formed during the initial disruption event contain up to $\approx 0.1M_{\odot}$, depending on the details of the encounter (merger vs. collision and assumed equation of state are the most important ones). The fraction that is bound to the central object will return on largely ballistic trajectories, which can be computed from the merger calculations. The differential distribution of mass with energy, $dm/d\epsilon$ is fairly flat, implying the fallback mass accretion rate follows a power law, $dM_{\rm fb}/dt \propto t^{-n}$, with $n\simeq 4/3-5/3$ (Rosswog 2007, Lee \& Ramirez-Ruiz 2007). The typical time scale for the bulk of this material, $M_{\rm fb}$ to return is $t_{\rm fb}\approx$~few seconds. The crucial point is that, since this material was dynamically ejected from the outer Lagrange points during disruption, it carries a significant  amount of specific angular momentum, and has a greater circularization radius than that which  made up the first accretion disk, of mass $M_{\rm disk}$. This time scale thus represents the delay required for a new accretion structure to form, not that on which any renewed activity it powers will be manifested. The latter will be the accretion time scale of the new disk, which will dominate the evolution of the system if $M_{\rm fb}\gg M_{\rm disk}$ at times greater than $t_{\rm fb}$ (Lee et al. 2009). Whether this is fulfilled or not will depend on the particular details of each encounter, and it is clear that not all events are capable of exhibiting such episodes, but time scales reaching minutes are attainable. 

\section{{\em Finale?}}

In a final gasp, the system may be capable of a last observable signal through radioactive decay of synthesized elements, either from material dynamically ejected during merger (Li \& Paczynski 1998) or from disk-driven winds (Lee et al. 2009, Metzger et al. 2009). Determining whether this is the case will require detailed calculations involving both the dynamics and the nuclear and thermodynamical properties of the outflows. The associated range of time scales in this case would run into days or even weeks after the main event. 

It is possible that compact object mergers are the source of (some?, most?) short-hard gamma-ray bursts, but they are, at least in our current understanding, incapable of being so for all such events.  Perhaps, as Maxim Lyutikov has argued in these proceedings, it is indeed D.C. al Fine.

\bigskip

{\bf Acknowledgments} It is a pleasure to thank the organizers for a stimulating meeting in a wonderful venue. Extended collaboration with Enrico Ramirez-Ruiz is warmly acknowledged. 



%
\section{References}
Balbus, S.A., Hawley, J.F. 1991, ApJ, 376, 214 \\
Baumgarte, T. W., Shapiro, S. L., Shibata, M. 2000, ApJ, 528, L28 \\
Burgay, M. et al. 2003, Nature, 426, 531 \\ 
Chevalier, R.A. 1989, ApJ, 346, 847 \\
Cook, G. B., Shapiro, S. L., Teukolsky, S. A. 1994, ApJ, 424, 823 \\
Cutler, C. Holz, D.E. 2009, Phys. Rev. D, 80, 104009 \\
Di Matteo, T., Perna, R., Narayan R. 2002, ApJ 579, 706 \\
Faber, J. A., Grandcl\'{e}ment, P., Rasio, F. A. 2002, Phys. Rev. Lett. 89, 231102 \\
Hulse, R.A. Taylor, J.H. 1975, ApJ, 195, L51 \\
Lai D., Rasio, F.A., Shapiro, S.L., 1993, ApJ, 406, L63 \\
Lattimer, J.M.,  Schramm, D.N. 1976, ApJ, 210, 549 \\
Lattimer, J.M., Prakash, M. 2007, Phys. Rep. 442, 109 \\
Lee, W.H. 2000, MNRAS, 318, 606 \\
Lee, W.H., Ramirez-Ruiz, E. 2007, New J. Phys., 9, 17 \\
Lee, W. H., Ramirez-Ruiz, E., Page, D. 2004 ApJ, 608, L5 \\
Lee, W. H., Ramirez-Ruiz, E., Page, D. 2005 ApJ, 632, 421 \\
Lee, W. H., Ramirez-Ruiz, E., Lopez-Camara, D. 2009, ApJ, 699, L93 \\
Li, Li-X., Paczynski, B. 1998, ApJ, 507, L59 \\
Lyutikov, M. 2006, New J. Phys., 8, 119 \\
Metzger, B.D., Quataert, E., Thompson, T.A. 2008, MNRAS, 385, 1455 \\
Metzger, B.D., Piro, A., Quataert, E. 2009, MNRAS, 396, 304 \\
Nakar, E. 2007, Phys. Rep. 442, 166 \\
Paczynski, B. 1986, ApJ, 308, L43 \\
Price, D., Rosswog, S. 2006, Science, 312, 719 \\
Rosswog, S. 2007, MNRAS, 376, L48 \\
Rosswog, S., Davies, M.B. 2002, MNRAS, 334, 481 \\
Setiawan, S., Ruffert, M. Janka, H.-Th. 2004, MNRAS, 352, 753 \\
Shakura, N.I., Sunyaev, R.A. 1973, A\&A, 24, 337 \\
Usov, V.V. 1992, Nature, 357, 472 \\

\end{document}